\documentclass[a4paper,12pt]{article}
\renewcommand{\baselinestretch}{2}
\usepackage{graphicx}%
\usepackage{graphics}
\usepackage{amsmath,amsfonts,amssymb}
\usepackage{subfigure}
\usepackage{overcite}
\usepackage{lscape}
\begin{document}

\section{Copyright}

Copyright 2009 American Institute of Physics. This article may be downloaded for personal use only. Any other use requires prior
permission of the author and the American Institute of Physics. The following article appeard in J.\ Chem.\ Phys.\, \textbf{130}, 174101 (2009)  and may be found at\\
 http://jcp.aip.org/resource/1/jcpsa6/v130/i17/p174101\_s1

\title{Accurate interaction energies at DFT level by means of an efficient dispersion correction}
\author{Alisa Krishtal$^{1}$ \and Kenno Vanommeslaeghe$^{2}$ \and Andr\'{a}s Olasz$^{3,4}$ \and   Tam\'as Veszpr\'emi$^{3}$ \and Christian Van Alsenoy$^{1}$,\and  and Paul Geerlings$^{4}$} 
\date{} 

\maketitle
\noindent $^1$Department of Chemistry, University of Antwerp, Universiteitsplein 1, B-2610 Ant\-werp, Belgium 
\noindent $^2$Department of Pharmaceutical Sciences, University of Maryland School of Pharmacy, 20 Penn St., HSF II-629, Baltimore, Maryland 21201, USA\\
\noindent $^3$Department of Inorganic Chemistry, Budapest University of Technology and Economics, Gell\`{e}rt t\'{e}r 4, H-1521 Budapest, Hungary\\
\noindent $^4$Algemene Chemie, Vrije Universiteit Brussel, Pleinlaan 2, B-1050 Brussels, Belgium

\vspace{2cm}
\begin{abstract} 
This paper presents an approach for obtaining accurate interaction energies at the DFT level for systems where dispersion interactions are important. This approach combines Becke and Johnson's [J. Chem. Phys. \textbf{127}, 154108 (2007)] method for the evaluation of dispersion energy corrections and a Hirshfeld method for partitioning of molecular polarizability tensors into atomic contributions. Due to the availability of atomic polarizability tensors, the method is extended to incorporate anisotropic contributions, which prove to be important for complexes of lower symmetry. The method is validated for a set of eighteen complexes, for which interaction energies were obtained with the B3LYP, PBE and TPSS functionals combined with the aug-cc-pVTZ basis set and compared with the values obtained at CCSD(T) level extrapolated to a complete basis set limit. It is shown that very good quality interaction energies can be obtained by the proposed method for each of the examined functionals, the overall performance of the TPSS functional being the best, which with a slope of 1.00 in the linear regression equation and a constant term of only 0.1 kcal/mol allows to obtain accurate interaction energies without any need of a damping function for complexes close to their exact equilibrium geometry.

\textbf{Keywords}: Interaction Energies, Dispersion, Hirshfeld, DFT, Basis Set Extrapolation. 
\end{abstract}
\newpage
\section{Introduction}

Since Density Functional Theory has been introduced into the world of computational chemistry, countless studies on ``large'' systems have been performed that would otherwise have been impossible due to the size of the systems examined.\cite{ref:Koch} A large portion of those studies involve biologically active molecules, their structure, reactivity, catalytic and binding properties.\cite{ref:HobzaPCCP} One of the most fundamental aspects examined in these studies are interaction energies. However, the use of DFT becomes problematic when energetics and related properties are examined for systems where dispersion interactions are important.\cite{ref:Hobza} Accurate description of interaction energies demands the use of levels of theory that include electron correlation, and although MP2 has started to become applicable to systems of relevant size in recent years\cite{ref:Versees,ref:Sinnokrot,ref:Mignon1,ref:Mignon2,ref:Mignon3}, the more accurate methods such as CCSD(T) are still far from reaching that stage. Therefore, the adjustment of DFT methods for a correct descriptions of dispersion interaction is nowdays a topic of an active research. Numerous examples can be found in the recent special issue of Phys. Chem. Chem. Phys. dedicated to stacking interactions.\cite{ref:HobzaPCCP}

In several studies adjustments were made to existing density functionals to improve their performance for non-covalent interactions. Xu et al.\cite{ref:Xu} designed an X3LYP extended functional, based on the well-known B3LYP functional, that improves the accuracy for Van der Waals complexes. Zhao and Truhlar\cite{ref:Zhao1,ref:Zhao2,ref:Zhao3} developed functionals based on simultaneously optimized exchange and correlation functionals. Rothlisberger et al.\ developed a dispersion-corrected DFT, where they augment the B3LYP functional with dispersion corrected atom-centered potentials (DCACPs).\cite{ref:Lilienfeld1,ref:Lilienfeld2,ref:Lin1,ref:Lin2}. Also several studies by Hirao and Lundqvist have been performed for developing special correlation functionals which take long-range dispersion interactions into account.\cite{ref:Dion,ref:Ikura,ref:Sato,ref:Kamiya}. The inclusion of empirical dispersion  coefficients\cite{ref:Grimme1,ref:Grimme2,ref:Grimme3,ref:Grimme4,ref:Grimme5,ref:Jurecka,ref:Cerny,ref:Chai}, as advocated by Grimme, Hobza and Head-Gordon, has booked some success as a cost-efficient method for examination of stacking phenomena in larger systems, since such applications are today not yet possible with the more elaborated methods mentioned above.

A different approach consists of calculating dispersion energies from dispersion coefficients. For instance, Van Gisbergen et al.\ derived van der Waals dispersion coefficients from frequency dependent polarizabilities using time dependent DFT.\cite{ref:VanGisbergen1,ref:VanGisbergen2} On the other hand, Becke and Johnson\cite{ref:BJ1,ref:BJ2,ref:BJ3,ref:BJ4,ref:BJ5,ref:BJ6} developed an approximation for the calculation of dispersion coefficients from exchange-hole dipole moments that allows to obtain dispersion energies at the DFT level in an easy and efficient fashion. However, the method of Becke and Johnson has two major drawbacks. First of all, Becke and Johnson use inappropriate values for the polarizabilities of atoms in the molecules, thus undermining the theoretical foundation of the method significantly. In their first publication they used empirical values for polarizabilities, while in their later works they approximated atomic polarizabilities by scaling free-atom polarizabilities by the Hirshfeld effective volume of the atom in the molecule. Such a rough approach for obtaining atom-in-molecule polarizabilities ignores many relevant effects, such as electron density reorganization due to the applied electric field. Second, Becke and Johnson make use of a damping function that strongly reduces the values of dispersion energy even at equilibrium geometries by an approximate factor of 2.

In our previous work\cite{ref:disp1} we showed that the use of intrinsic polarizabilities, obtained from the Hirshfeld method\cite{ref:Hirshfeld,ref:Rousseau,ref:polar}, improves the dispersion energies obtained from the dispersion coefficients of Becke and Johnson significantly. As a result, not only the dubious character of the atomic polarizabilities used in the method is eliminated, but also realistic dispersion energies are obtained at equilibrium geometry without the need for a damping function.

In this work we develop our all-Hirshfeld approach further in three separate ways. First of all, we extend the methodology for reproducing high-level interaction energies at the DFT level, instead of comparing pure dispersion energies. This method is applied using three functionals of a different nature, namely the hybrid functional B3LYP,\cite{ref:B3LYP} the GGA functional PBE\cite{ref:PBE1,ref:PBE2} and the meta-GGA functional TPSS,\cite{ref:TPSS} none of them including non-local correlation, in order to test its universal character. Second, in order to improve the accuracy for complexes of reduced symmetry, we introduce anisotropy for the derivation of the dispersion coefficients. Finally, we introduce the iterative Hirshfeld method (Hirshfeld-I)\cite{ref:H-I} into the calculation of the dispersion coefficients. The Hirshfeld-I method, recently developed by Bultinck et al.\cite{ref:H-I}, brings several fundamental improvements to the classic Hirshfeld weight function and the resulting partitioned properties such as charges, dipole moments and polarizabilities. By its iterative nature the method eliminates the somewhat arbitrary  nature of the weight function and can also be applied to charged systems. It has recently been shown by some of the authors that not only the charges obtained with Hirshfeld-I are more in line with the oxidation state of the atoms in the molecules, but also the intrinsic polarizabilities are more adequate.\cite{ref:SulfonicAcids} For the validation of our modified approach a set of 18 complexes is examined. To ensure the quality of the high level geometries and interaction energies, all the complexes were optimized using the same methodology, namely CCSD(T)/aug-cc-pVTZ and a frozen monomer approach.

\section{Method}
The dispersion corrections used in this work are based on the model developed by Becke and Johnson\cite{ref:BJ1,ref:BJ2,ref:BJ3,ref:BJ4,ref:BJ5}, wherein explicit expressions for the dispersion coefficients $C_{6}$, $C_{8}$ and $C_{10}$ were derived from the instantaneous dipole moment created by an electron and its corresponding Fermi hole. The dispersion energy between two nonoverlapping systems A and B at a distance R from each other is given by
\begin{equation}
E_{disp}=-\left( \frac{C_{6}}{R^{6}}+\frac{C_{8}}{R^{8}}+\frac{C_{10}}{R^{10}}\right) .\label{eq:disp-energy}
\end{equation}
According to Becke and Johnson's model the coefficients in eq.\ (\ref{eq:disp-energy}) can be obtained from the polarizabilities $\alpha$ of the systems and the expectation values of the square of their dipole moments $\left<M^{2}_{1}\right>$, quadrupole moments $\left<M^{2}_{2}\right>$ and octopole moments $\left<M^{2}_{3}\right>$
\begin{equation}
C_{6}=\frac{\alpha_{A}\alpha_{B}\left<M^{2}_{1}\right>_{A}\left<M^{2}_{1}\right>_{B}}{\alpha_{A}\left<M^{2}_{1}\right>_{B}+\alpha_{B}\left<M^{2}_{1}\right>_{A}}, \label{eq:c6}
\end{equation}
\begin{equation}
C_{8}=\frac{3}{2}\frac{\alpha_{A}\alpha_{B}\left(\left<M^{2}_{1}\right>_{A}\left<M^{2}_{2}\right>_{B}+\left<M^{2}_{2}\right>_{A}\left<M^{2}_{1}\right>_{B}\right)}%
{\alpha_{A}\left<M^{2}_{1}\right>_{B}+\alpha_{B}\left<M^{2}_{1}\right>_{A}}, \label{eq:c8}
\end{equation}
\begin{eqnarray}
C_{10}&=&2\frac{\alpha_{A}\alpha_{B}\left(\left<M^{2}_{1}\right>_{A}\left<M^{2}_{3}\right>_{B}+\left<M^{2}_{3}\right>_{A}\left<M^{2}_{1}\right>_{B}\right)}{\alpha_{A}\left<M^{2}_{1}\right>_{B}+\alpha_{B}\left<M^{2}_{1}\right>_{A}} \notag\ \\
 &+&\frac{21}{5}\frac{\alpha_{A}\alpha_{B}\left(\left<M^{2}_{2}\right>_{A}\left<M^{2}_{2}\right>_{B}+\left<M^{2}_{2}\right>_{A}\left<M^{2}_{2}\right>_{B}\right)}
{\alpha_{A}\left<M^{2}_{1}\right>_{B}+\alpha_{B}\left<M^{2}_{1}\right>_{A}}. \label{eq:c10}
\end{eqnarray}
The expectation values of the square of a multipole $M_{l}$ are approximated by
\begin{equation}
\left<M^{2}_{l}\right>=\sum_{\sigma}\int\rho_{\sigma}(\mathbf{r})\left[\mathbf{r}^{l}-(\mathbf{r}-\mathbf{d}_{X\sigma})^{l}\right]^{2}d^{3}\mathbf{r}, \label{eq:expec}
\end{equation}
with $\sigma$ representing the spin of the electron.

It has also been shown by Becke and Johnson that when the systems $A$ and $B$ contain more than one atom, the dispersion energies obtained from coefficients in equations (\ref{eq:c6}-\ref{eq:c10}) can be decomposed into pair-wise atom-atom interactions between the atoms in the two systems. For example, the interaction energy obtained from the $C_{6}$ term can be decomposed into pair-wise contributions as
\begin{equation}
E_{disp,6}=\sum^{A}_{a}\sum^{B}_{b}\frac{C_{6,ab}}{R^{6}_{ab}}
\end{equation}
with
\begin{equation}
C_{6,ab}=\frac{\alpha_{a}\alpha_{b}\left<M^{2}_{1}\right>_{a}\left<M^{2}_{1}\right>_{b}}{\alpha_{a}\left<M^{2}_{1}\right>_{b}+\alpha_{b}\left<M^{2}_{1}\right>_{a}}
\end{equation}
In the expressions for these interatomic coefficients the polarizabilities and expectation values of the squares of the multipole moments of the systems are replaced by atomic polarizabilities and expectation values of the squares of the atomic multipole moments. In our previous work\cite{ref:disp1} we suggested an all-round Hirshfeld approach, where the Hirshfeld atomic multipole moments\cite{ref:Hirshfeld,ref:Rousseau,ref:DeProft} and Hirshfeld atomic intrinsic polarizabilities\cite{ref:polar} are employed. 

The Hirshfeld method allows to partition properties into atomic contributions by means of a weight function. The weight of each atom is determined by the density of the corresponding free spherical atom, normalized by the sum of all the free atomic densities of the atoms in the molecule
\begin{equation}
\omega_{A}(\mathbf{r})=\frac{\rho_{A}^{free}(\mathbf{r})}{\sum_{B}\rho_{B}^{free}(\mathbf{r})}. \label{eq:hirshfeld-classic}
\end{equation}
The elements of the intrinsic atomic polarizability tensor are then defined by\cite{ref:polar}
\begin{equation}
\alpha^{A}_{ij}=\int i_{A}\, \omega_{A}(\mathbf{r}) \rho^{(j)}(\mathbf{r}) d\mathbf{r}, \label{eq:def-polarint}
\end{equation}
where $i$ and $j$ represent the Cartesian directions $x,y$ or $z$ and $\rho^{(j)}(\mathbf{r})$ denotes the first order density perturbed by an electric field applied in direction $j$.
Since the polarizability is not a straightforwardly additive property, the total polarizability of the molecule cannot be reconstructed from the intrinsic polarizabilities alone, but a charge transfer term must be added. However, in the present work we will only consider the intrinsic polarizabilities.

In this work the method is further extended by introducing the improved, iterative Hirshfeld approach (Hirshfeld-I), recently developed by Bultinck et al.\cite{ref:H-I} to the calculation of dispersion coefficients. The Hirshfeld-I method differs from the classic Hirshfeld method (Hirshfeld-C) by the definition of the atomic weight function. Whereas in Hirshfeld-C the weight function is predefined by the atomic densities of the free spherical atoms, in Hirshfeld-I the weight function is iterated until self consistency and therefore loses its somewhat arbitrary character. The Hirshfeld-C weight function is used as a first guess, and in each consecutive iteration the new weight function is constructed from the atomic densities obtained from the weight function of the previous iteration
\begin{equation}
\omega^{n}_{A}(\mathbf{r})=\frac{\rho_{A}^{n-1}(\mathbf{r})}{\sum_{B}\rho_{B}^{n-1}(\mathbf{r})}. \label{eq:hirshfeld-iterative}
\end{equation}
The process is repeated untill the weight functions of two subsequent iterations are identical.

The use of atomic intrinsic polarizabilites, which are obtained in the form of an atomic polarizability tensor, allows to introduce anisotropy into the model described above.

%Hier komt deel Paul Geerlings

Going back to the classical paper by Buckingham\cite{ref:Buckingham,ref:Buckingham2}, standard second order Rayleigh-Schr\"{o}dinger perturbation theory for long range intermolecular forces was shown to yield the following general expression for the $R^{-6}$ contribution to the dispersion energy for two molecules or two atoms $a$ and $b$
\begin{equation}
E_{ab,disp}(R^{-6})=-\frac{U_{a}U_{b}}{4(U_{a}+U_{b})}\sum_{i}\sum_{j}\sum_{k}\sum_{l}T_{2,ij}T_{2,kl}
\alpha^{(a)}_{ik}\alpha^{(b)}_{jl} \label{eq:Eabdisp}
\end{equation}
where $\alpha^{(a)}$ and $\alpha^{(b)}$ are the (dipole) polarizability tensors of the interacting systems and the elements of the $T_{2}$ tensor are defined as
\begin{equation}
T_{2,ij}=\nabla_{i}\nabla_{j}R^{-1}_{ab}
\end{equation}
where $R_{ab}$ is the intermolecular distance and $i$ and $j$ stand for the Cartesian coordinates $x,y,z$.
This expression was obtained using an Uns\"{o}ld/London type of approximation\cite{ref:Unsold,ref:London} by simplifying the energy denominator in the second order perturbation theory expression by introducing an average excitation energy (or ionization energy), for $a$ and $b$, namely $U_{a}$ and $U_{b}$.
Equation (\ref{eq:Eabdisp}) can  then be rewritten as\cite{ref:Pullman}
\begin{equation}
E_{ab,disp}(R^{-6})=-\frac{U_{a}U_{b}}{4(U_{a}+U_{b})}\operatorname{Tr}{(\stackrel{\leftrightarrow}{\Pi} \stackrel{\leftrightarrow}{\alpha}^{(a)} \stackrel{\leftrightarrow}{\Pi} \stackrel{\leftrightarrow}{\alpha}^{(b)})} \label{eq:Eabdisp2}
\end{equation}
where the $\stackrel{\leftrightarrow}{\Pi}$ and $\stackrel{\leftrightarrow}{\alpha}$ tensors involve the elements $T_{2,ij}$ and $\alpha_{ij}$.
It is easily shown that in the case of \emph{isotropic} tensors equation (\ref{eq:Eabdisp2}) reduces to
\begin{equation}
E_{ab,disp}(R^{-6})=-\frac{1}{4}\frac{1}{R^{6}_{ab}}\frac{U_{a}U_{b}}{U_{a}+U_{b}}\alpha^{(a)}\alpha^{(b)}\operatorname{Tr}{(\stackrel{\leftrightarrow}{\Pi}^{2})}
\end{equation}
where $\alpha^{(a)}$ and $\alpha^{(b)}$ are now equal to the diagonal elements of $\stackrel{\leftrightarrow}{\alpha}^{(a)}$ and $\stackrel{\leftrightarrow}{\alpha}^{(b)}$ (or one third of their traces), respectively.
As it is easily shown that $\operatorname{Tr}{(\stackrel{\leftrightarrow}{\Pi}^{2})}=6$, eq.\ (\ref{eq:Eabdisp2}) finally reduces to
\begin{equation}
E_{ab,disp}(R^{-6})=-\frac{3}{2}\frac{1}{R^{6}_{ab}}\frac{U_{a}U_{b}}{U_{a}+U_{b}}\alpha^{(a)}\alpha^{(b)}
\end{equation}
whereby the standard London dispersion formula is recovered. Concentrating now on a pairwise atom-atom interaction scheme, where in (\ref{eq:Eabdisp2}) $a$ and $b$
refer to isolated atoms or atoms-in-molecules, and replacing in (\ref{eq:Eabdisp2}), in the spirit of Becke and Johnson's treatment\cite{ref:BJ1} as also adopted in our previous work, the average excitation energies $U_{a}$ and $U_{b}$ by expressions of the type $2\left<M_{1}^{2}\right>/3\alpha$, where $\alpha$ is the isotropic polarizability of the atom in the molecule, we arrive at
\begin{equation}
E_{disp,ab}^{aniso}(R^{-6})=-\frac{1}{6}\frac{\left<M^{2}_{1}\right>_{(a)}\left<M^{2}_{1}\right>_{(b)}}{\alpha_{b}\left<M^{2}_{1}\right>_{(a)}+\alpha_{a}\left<M^{2}_{1}\right>_{(b)}}\operatorname{Tr}{(\stackrel{\leftrightarrow}{\Pi}\stackrel{\leftrightarrow}{\alpha}^{(a)}\stackrel{\leftrightarrow}{\Pi}\stackrel{\leftrightarrow}{\alpha}^{b}}). \label{eq:Eabdisp3}
\end{equation} 
In the isotropic case the equation reduces to
\begin{equation}
E_{disp,ab}(R^{-6})=-\frac{\left<M^{2}_{1}\right>_{(a)}\left<M^{2}_{1}\right>_{(b)}}{\alpha_{b}\left<M^{2}_{1}\right>_{(a)}+\alpha_{a}\left<M^{2}_{1}\right>_{(b)}}\alpha_{a}\alpha_{b}\frac{1}{R^{6}_{ab}}
\end{equation}
which is the expression for the $R^{-6}$ term used in our previous work.\cite{ref:disp1}

Equation (\ref{eq:Eabdisp3}) was implemented in the \textsc{atdisp}\cite{ref:disp1} program. Note that anistropy corrections to the $R^{-8}$ and $R^{-10}$ terms could be treated in a similar way necessitating, however, the evaluation of terms involving quadrupole and mixed dipole-quadrupole polarizabilities.   As these contributions are expected to be smaller and  lend themselves less easily to an interpolation into the framework of eq.\ (\ref{eq:Eabdisp3}),  we only consider in this paper the expression of an anisotropy corrected $C_{6}$ term, optionally combined with isotropic $C_{8}$ and $C_{10}$ terms. On the whole, anisotropy corrections to dispersion coefficients were relatively seldom studied in the literature.\cite{ref:Rijks}.

\section{Computational Details}

The main goal of this work is to reproduce accurate interaction energies obtained from high level calculations by adding dispersion energy corrections to interaction energies obtained at the DFT level. For this purpose, a set of eighteen different complexes was examined. In order to ensure that the benchmark set is of good and consistent quality, individual geometries of the different complexes have been optimized as follows. First the geometries of the monomers have been fully optimized at the CCSD(T)/aug-cc-pVTZ level with tight convergence criteria. Subsequently, the geometries of the complexes were optimized at the same level keeping the internal geometries of the monomers frozen. For most of the complexes this meant optimizing only one parameter, namely the distance between the two monomers. For a few others the lateral displacement of the two monomers has been taken into account. Symmetries of the complexes with the lowest reported energies were taken from the literature.\cite{ref:geom-atom,ref:geom-hen2,ref:geom-hefcl,ref:Zhao1,ref:BJ2,ref:geom-sih4ch4,ref:geom-co2co2,ref:geom-ocsocs}
Once the equilibrium geometries were obtained, the interaction energies at CCSD(T) level and at DFT level using the B3LYP, PBE and TPSS functionals were calculated, taking into account the counterpoise BSSE correction\cite{ref:bsse-Boys,ref:bsse-Simon}. Since the geometry of the monomers was kept unchanged in the dimers, the interaction energies are given by
\begin{equation}
E^{CCSD(T)}_{inter}=E^{CCSD(T)}_{AB}-\left[E^{CCSD(T)}_{A}\right]^{*}-\left[E^{CCSD(T)}_{B}\right]^{*} \label{eq:inter-ccsd}
\end{equation}
\begin{equation}
E^{DFT}_{inter}=E^{DFT}_{AB}-\left[E^{DFT}_{A}\right]^{*}-\left[E^{DFT}_{B}\right]^{*} \label{eq:inter-b3lyp}
\end{equation}
The stars in eq.\ (\ref{eq:inter-ccsd}) and (\ref{eq:inter-b3lyp}) denote that the energies were obtained using all the basis functions of the dimer.

To ensure high quality reference data the values for the interaction energy at the CCSD(T) level were extrapolated to the complete basis set limit (CBS) using the following focal point analysis. To obtain the BSSE-corrected interaction energy one needs to extrapolate the energies of the dimers and the complexes
\begin{equation}
E^{CCSD(T)/CBS}_{inter}=E^{CCSD(T)/CBS}_{AB}-\left[E^{CCSD(T)/CBS}_{A}\right]^{*}-\left[E^{CCSD(T)/CBS}_{B}\right]^{*} \label{eq:inter-ccsd-cbs}
\end{equation}
where the total energy of each entity is a sum of the Hartree-Fock energy and the correlation energy
\begin{equation}
E^{CCSD(T)/CBS}_{tot}=E_{tot}^{HF/CBS}+E^{CCSD(T)/CBS}_{corr}
\end{equation}
Since it has been shown by Sinnokrot and Sherrill\cite{ref:Sinnokrot2} that the correlation energies at the MP2 and CCSD(T) levels converge very similarly with the size of the basis set, the difference between the two remaining constant for the set of the aug-cc-pVXZ basis sets upon the increase of X, it is sufficient to extrapolate the correlation energy at the MP2 level to the basis set limit and add the correlation energy value between the two methods obtained for a smaller basis set:
\begin{equation}
E^{CCSD(T)/CBS}_{corr}=E^{MP\textit{2}/CBS}_{corr}+[E^{CCSD(T)/aug-cc-pVTZ}_{corr}-E^{MP\textit{2}/aug-cc-pVTZ}_{corr}]
\end{equation}
The MP2 correlation energy at the complete basis set limit was obtained using Helgaker's linear extrapolation formula\cite{ref:Helgaker1,ref:Helgaker2}
\begin{equation}
E^{MP\textit{2}/CBS}_{corr}=\frac{X^{3}E^{MP\textit{2}/aug-cc-pVXZ}_{corr}-Y^{3}E^{MP\textit{2}/aug-cc-pVYZ}_{corr}}{X^{3}-Y^{3}}
\end{equation}
where we chose X=5 and Y=4 for all the complexes.
Finally, since the Hartree-Fock total energy converges towards the complete basis set limit fast and monotonous\-ly,\cite{ref:Helgaker1} the energies at the HF/aug-cc-pV6Z level were used to estimate $E^{HF/CBS}_{tot}$.

In the following step the interaction energy at the DFT level, obtained with the B3LYP, PBE and TPSS functionals, was corrected for dispersion
\begin{equation}
E^{corr. DFT}_{inter}=E^{DFT}_{inter}+E^{DFT}_{disp}, \label{eq:ener-corr}
\end{equation}
where the dispersion interaction correction was calculated in four ways of increasing complexity. The first two approaches, which neglect anistropy, yield the following equations:
\begin{itemize}
\item  Isotropic $C_{6}$ term  only
\begin{equation}
E^{DFT}_{disp,C^{iso}_{6}}=-\sum_{a}^{A}\sum_{b}^{B}\frac{C^{iso}_{6,ab}}{R^{6}_{ab}} \label{eq:disp-c6}
\end{equation}
\item  Three isotropic terms 
\begin{equation}
E^{DFT}_{disp,full}=-\sum_{a}^{A}\sum_{b}^{B}\left(\frac{C^{iso}_{6,ab}}{R^{6}_{ab}}+\frac{C^{iso}_{8,ab}}{R^{8}_{ab}}+\frac{C^{iso}_{10,ab}}{R^{10}_{ab}}\right) \label{eq:disp-c6c8c10}
\end{equation}
\end{itemize}
In a subsequent step anisotropy is introduced. In order to reach an optimal quality cost ratio only the $C_{6}$ term was corrected for anistropy as it may be expected that the main contribution to anisotropy will essentially be due to the $C_{6}$ term. The counterpart of eq.\ (\ref{eq:disp-c6}) may be written as
\begin{itemize}
\item  Anisotropic $C_{6}$ term only
\begin{equation}
E^{DFT}_{disp,C^{aniso}_{6}}=-\sum_{a}^{A}\sum_{b}^{B}E_{disp,ab}^{aniso}(R^{-6}) \label{eq:disp-anisoc6}
\end{equation}
\item A compilation  of the anisotropic $C_{6}$ term and isotropic $C_{8}$ and $C_{10}$ terms finally gives
\begin{equation}
E^{DFT}_{disp,mixed}=-\sum_{a}^{A}\sum_{b}^{B}\left(E_{disp,ab}^{aniso}(R^{-6})+\frac{C^{iso}_{8,ab}}{R^{8}_{ab}}+\frac{C^{iso}_{10,ab}}{R^{10}_{ab}}\right) .\label{eq:anisodisp-energy}
\end{equation}
\end{itemize}
%The isotropic coefficients used in eq. (\ref{eq:disp-c6}) to (\ref{eq:anisodisp-energy}) were obtained by both Hirshfeld-C and Hirshfeld-I method.
The coefficients used in eq.\ (\ref{eq:disp-c6}) to (\ref{eq:anisodisp-energy}) were obtained by the Hirshfeld-I method. Note that no damping function has been used in these equation, since we are interested here in dispersion energies of complexes at their equilibrium geometry, where the interatomic distances are relatively large. 

The geometries and polarizabilities of the different molecules were calculated using the \textsc{gaussian03}\cite{Gaussian} program. The values of atomic intrinsic polarizabilities and expectation values of the squared atomic multipole moments were obtained using the \textsc{stock}\cite{ref:Rousseau} program. Finally the different dispersion coefficients and the corresponding dispersion energy corrections were calculated using the \textsc{atdisp}\cite{ref:disp1} program.

\section{Results and Discussion}

Table 1 lists the interaction energies obtained with the CCSD(T)/CBS, B3LYP/aug-cc-pVTZ, PBE/aug-cc-pVTZ and TPSS/aug-cc-pVTZ methods. The high level interaction energies vary from -0.02 to -1.76 kcal/mol, whereas the interaction energies obtained at the DFT level seem to be dependent on the choice of functional. For the B3LYP functional all interaction energy values are repulsive, as expected at the DFT level where non-local dispersion interactions are not included in the exchange-correlation functional. However, PBE produces negative interaction energies for all of the examined complexes while the values obtained with the TPSS functional are partially negative. Since the attraction for most of the examined complexes can be attributed only to dispersion energy, the negative interaction values at the DFT level are not physically justified, B3LYP thus being in this aspect the most reliable of the three functionals examined. 

Tables 2 to 4  give the dispersion corrected post-DFT interaction energies obtained by the $C_6$-isotropic (eq.\ \ref{eq:disp-c6}), full isotropic (eq.\ \ref{eq:disp-c6c8c10}), $C_6$ anisotropic (eq.\ \ref{eq:disp-anisoc6}) and mixed (eq.\ \ref{eq:anisodisp-energy}) models  for the B3LYP, PBE and TPSS functionals, respectively. In contrast to our previous work on dispersion energies,\cite{ref:disp1} where the classic version of the Hirshfeld method was utilized, the dispersion corrections are obtained here with the Hirshfeld-I method. The difference between the two versions of the method is only of importance for the larger complexes, where the monomers contain  more than one atom, which are unidentical. In that case the atomic weight functions which determine the distribution of the electronic density among the atoms are different, and as a result also the charges and other properties of the atoms are different. It has been shown in a previous study by some of the authors\cite{ref:SulfonicAcids} that atomic polarizabilities obtained with the Hirshfeld-I method are of better quality, therefore also leading to more reliable dispersion coefficients. The two last lines in Tables 2 to 4 also show the correlation coefficient between the post-DFT interaction energies and the high level interaction energies listed in Table 1 and the standard error of the linear regression.

From Table 2 it appears that for the B3LYP functional, the addition of a dispersion interaction correction based on the $C_{6}$ dispersion coefficient reduces most of the interaction energies to negative values, although for five complexes the interaction energies remain positive. Taking into account the remaining $C_{8}$ and $C_{10}$ coefficients further reduces all post-B3LYP interaction energies to values similar to the high-level interaction energies. Figure 1 depicts the correlation between the post-B3LYP interaction energy values, obtained using the isotropic and anisotropic $C_{6}$ dispersion coefficients, and the high level values. Although the correlation is poor, being only $R=0.7064$ and $R=0.8098$ for the isotropic and anisotropic models, respectively, the effect of the addition of anisotropy can be seen very clearly in this Figure. Obviously, the anisotropic contributions are only of significance for the complexes where the monomers are not spherically symmetric, so the values for the smaller complexes remain the same. While for the isotropic model the values for the CO$_{2}$-CO$_{2}$ and C$_{2}$H$_{4}$-C$_{2}$H$_{4}$ complexes are situated far from the trend line, introduction of the anisotropy to the $C_{6}$ coefficient lowers the value for the C$_{2}$H$_{4}$-C$_{2}$H$_{4}$, although the CO$_{2}$-CO$_{2}$ complex remains an outlier.  One can conclude that the large standard error (listed in Table 2) and the very low value of the slope (shown in Figure 1), indicate that the $C_{6}$-based models are insufficient by far for reproducing accurate post-DFT interaction energies and that the higher coefficients are indispensable.

Figure 2 depicts the correlation between the high level interaction energy values and the post-B3LYP interaciton energy values for the full isotropic and mixed models. 
As was mentioned in the method section, the derivation of anisotropic dispersion coefficients higher than $C_{6}$ becomes complicated as extra terms appear in the equations and the model loses its simplicity, which is one of its major advantages. Therefore a mixed model is examined here, which on the one hand combines the improvement achieved in the $C_{6}$ coefficient by introducing anisotropy and on the other hand uses isotropic $C_{8}$ and C$_{10}$ coefficients, which are vital for reproduction of accurate interaction energies. The mixed model performs better than the full isotropic method, both of them performing significantly better than the $C_{6}$-based models, having a correlation coefficient of 0.94 and 0.95 for the former and the latter, respectively. Also the value for the CO$_{2}$-CO$_{2}$ complex improves considerably by the addition of the higher dispersion coefficients. The difference between the two models is the largest for the four complexes with the highest interaction energy values, namely CO$_{2}$-CO$_{2}$, OCS-OCS, C$_{2}$H$_{2}$-C$_{2}$H$_{2}$ and C$_{2}$H$_{4}$-C$_{2}$H$_{4}$. As can be seen in Figure 2, when only isotropic contributions are taken into account, the order of interaction energies for those three complexes is not correctly reproduced: while CCSD(T) interaction energies follow the order (in absolute values) OCS-OCS $>$ C$_{2}$H$_{4}$-C$_{2}$H$_{4}$ $>$ C$_{2}$H$_{2}$-C$_{2}$H$_{2}$ $>$ CO$_{2}$-CO$_{2}$, the interaction energies obtained with the full isotropic model follow the order C$_{2}$H$_{4}$-C$_{2}$H$_{4}$ $>$ OCS-OCS $>$ C$_{2}$H$_{2}$-C$_{2}$H$_{2}$ $>$ CO$_{2}$-CO$_{2}$. However, once anisotropy is introduced in the mixed model, the correct order is restored. The reason for this difference is the lower symmetry of the complexes, where the monomers are laterally shifted with respect to each other and anisotropy becomes more important. Therefore one can expect the linear regression parameters to improve further for the mixed model when more complexes of lower symmetry are taken into account. 

Although the correlation achieved for the full isotropic and mixed models is satisfying, the linear regression parameters are not optimal yet. With a slope of 0.82 in the full isotropic model and 0.83 in the mixed model, both models underestimate the interaction energy by an approximate 20\%. The PBE functional seems to suffer from an opposite problem, as can be seen from the values listed in Table 3. The interaction energies obtained by the full and mixed models are overestimated by no less than 40\%, as can also be seen from the linear regression coefficients in Figure 3. The source of the overestimation of the values must be sought in the pure DFT interaction energies obtained with this functional. As can be seen in Table 1, the PBE functional produces all negative interaction energies even though dispersion is not included in the functional. Since the dispersion energy correction is added to the original pure DFT interaction value, the spurious potential well produced by the PBE functional for the examined complexes causes a serious overestimation of the interaction energies. On the other hand, the correlation coefficients for the full isotropic and mixed models are very high, being more than 0.99, while the standard error is reduced almost half in size in comparison to the B3LYP functional. One can therefore conclude that our method for the calculation of dispersion energy corrections performs surprisingly well for the PBE functional, but for accurate interaction energies the values must be scaled by a factor of 0.71. It must also be mentioned that the effect of anisotropy on the interaction energy values is analogous for the PBE functional: the replacement of the isotropic $C_{6}$ coefficient by an anisotropic one in the mixed model restores the correct order in interaction energy values for the four largest complexes. The addition of the higher order coefficients $C_{8}$ and $C_{10}$ is here also of importance. Although the correlation coefficients are quite high for the $C_{6}$-based models, the values are less reliable. For example, the interaction value for the C$_{2}$H$_{2}$-C$_{2}$H$_{2}$ is an evident outlier in Table 3 for those two models.

The post-TPSS interaction energies  are listed in Table 4 for the four different models and depicted in Figure 4 for the full isotrpic and mixed models. This functional seems to perform very well, the correlation for the full isotropic and mixed models being 0.98 and the standard error on the linear regression being almost as low as for the PBE functional. The main strength of this functional is the perfect slope of 1.00 for the mixed model, enabling us to reproduce accurate interaction energies without the need for up- or down-scaling, as is the case for the B3LYP and PBE functionals. Amongst the larger complexes only the interaction energy value of the ethene dimer appears to be problematic, being overestimated by 0.4 kcal/mol. As was also the case with the PBE functional, the source for this overestimation may lay in the negative pure DFT interaction energy value of this complex (-0.548 kcal/mol). It seems that also here our method for obtaining dispersion energies is performing very well, but care must be taken if the potential energy surface produced by the functional is incorrect. 

\section{Concluding Remarks}

In our previous work\cite{ref:disp1} it was shown that the combination of Becke and Johnson's exchange dipole moment approach and our Hirshfeld-type partitioning scheme for molecular polarizabilities lead to a simple, accurate and inexpensive approach for evaluation of dispersion energies of dimers. This approach was further developed to reproduce the chemically more interesting interaction energies at the CCSD(T) level with simple dispersion energy corrected DFT interaction energies. Three functionals different by nature were examined here to test the robustness of our proposed method, namely B3LYP, PBE and TPSS. Two major conclusions can be drawn from the present results. First of all, the inclusion of anisotropy in the evaluation of the $C_{6}$ coefficient leads to an improvement of the corresponding dispersion energies. This could be seen especially for the dispersion energy values of the four complexes with lower symmetry, namely  CO$_{2}$-CO$_{2}$, OCS-OCS, C$_{2}$H$_{2}$-C$_{2}$H$_{2}$ and C$_{2}$H$_{4}$-C$_{2}$H$_{4}$, where the mixed model produced significant improvements. Secondly, our method performs well for different functionals regardless of their nature and might therefore be universally applicable to different DFT functionals. However, since the final result for the interaction energy is not only dependent on our dispersion energy correction but also on the pure DFT interaction energy, care must be taken in the choice of the functional. From the three functionals examined here TPSS seems to perform the best, producing accurate interaction energies for most of the complexes. However, for complexes where the TPSS functional produces questionable pure DFT interaction energies, such as for example was the case for the C$_{2}$H$_{2}$-C$_{2}$H$_{2}$ complex, the B3LYP functional is a better choice.

\section{Acknowledgments}
A.K.\ and C.V.A.\ acknowledge the Flemish FWO for research grant nr.\ G.0629.06. We gratefully acknowledge the University of Antwerp for the access to the university's CalcUA supercomputer cluster. P.G. acknowledges FWO and VUB for continuous support to his group. The authors thank the the referees for their helpful suggestions.

%\section{References}

\renewcommand{\baselinestretch}{1}
\newpage

%\begin{landscape}
\section{Tables}
\begin{table}[ht!]
	\centering
		\begin{tabular}{lcccc}
		\hline
		Complex & CCSD(T) & B3LYP & PBE & TPSS \\
		\hline
He-He	&	-0.020 &	0.041 &	-0.056 &	-0.043	\\
He-Ne	&	-0.037 &	0.040 &	-0.083 &	-0.058	\\
He-Ar	&	-0.056 &	0.072 &	-0.078 &	-0.047	\\
Ne-Ne	&	-0.071 &	0.045 &	-0.108 &	-0.066	\\
Ne-Ar	&	-0.119 &	0.021 &	-0.123 &	-0.056	\\
Ar-Ar	&	-0.272 &	0.177 &	-0.120 &	0.017 	\\
L-He-N2	&	-0.043 &	0.088 &	-0.073 &	-0.033	\\
T-He-N2	&	-0.061 &	0.109 &	-0.075 &	-0.027	\\
He-FCl	&	-0.096 &	0.072 &	-0.116 &	-0.046	\\
FCl-He	&	--0.126 &	0.048 &	-0.220 &	-0.079	\\
Ne-CH$_{4}$	&	-0.175 &	0.066 &	-0.123 &	0.022	\\
CH$_{4}$-C$_{2}$H$_{4}$	&	-0.449 &	0.428 &	-0.101 &	0.158	\\
CH$_{4}$-CH$_{4}$	&	-0.537 &	0.490 &	-0.025 &	0.273	\\
SiH$_{4}$-CH$_{4}$	&	-0.824 &	0.549 &	-0.130 &	0.312	\\
C$_{2}$H$_{2}$-C$_{2}$H$_{2}$	&	-1.403 &	0.153 &	-0.928 &	-0.548	\\
CO$_{2}$-CO$_{2}$	&	-1.476 &	0.757 &	-0.388 &	0.133	\\
OCS-OCS	&	-1.761 &	0.761 &	-0.083 &	0.572	\\
C$_{2}$H$_{4}$-C$_{2}$H$_{4}$	&	-1.493 &	0.544 &	-0.284 &	0.431	\\
\hline		
\end{tabular}
\caption[]{Interaction energies calculated with CCSD(T)/CBS, B3LYP/aug-cc-pVTZ, PBE/aug-cc-pVTZ and TPSS/aug-cc-pVTZ methods. All values are in kcal/mol.}
\end{table}

\newpage
%Table 2 B3LYP+D
\begin{table}[ht!]
	\centering
		\begin{tabular}{lcccc}
		\hline
		Complex & C$_{6}^{iso}$&C$_{6}^{aniso}$&	full&	mixed \\
		\hline
He-He	&	0.01	&	0.01	&	-0.01	&	-0.01	\\
He-Ne	&	-0.01	&	-0.01	&	-0.04	&	-0.04	\\
He-Ar	&0.01	&	0.01	&	-0.04	&	-0.04	\\
Ne-Ne	&-0.03	&	-0.03	&	-0.08	&	-0.08	\\
Ne-Ar	&-0.09	&	-0.09	&	-0.19	&	-0.19	\\
Ar-Ar	&-0.08	&	-0.08	&	-0.34	&	-0.34	\\
L-He-N2	& 0.02	&	0.03	&	-0.02	&	-0.01	\\
T-He-N2	&0.02	&	0.01	&	-0.05	&	-0.05	\\
He-FCl	& -0.04	&	-0.04	&	-0.13	&	-0.13	\\
FCl-He	& -0.04	&	-0.04	&	-0.09	&	-0.09	\\
Ne-CH$_{4}$	& -0.15	&	-0.15	&	-0.28	&	-0.28	\\
CH$_{4}$-C$_{2}$H$_{4}$	& 0.06	&	0.03	&	-0.11	&	-0.14	\\
CH$_{4}$-CH$_{4}$	&	-0.06	&	-0.06	&	-0.30	&	-0.30	\\
SiH$_{4}$-CH$_{4}$ & -0.21	&	-0.19	&	-0.62	&	-0.61	\\
C$_{2}$H$_{2}$-C$_{2}$H$_{2}$	& -0.50	&	-0.48	&	-1.16	&	-1.15	\\
CO$_{2}$-CO$_{2}$	&	0.03	&	-0.04	&	-0.75	&	-0.81	\\
OCS-OCS	& -0.33	&	-0.47	&	-1.56	&	-1.71	\\
C$_{2}$H$_{4}$-C$_{2}$H$_{4}$	& -0.69	&	-0.55	&	-1.64	&	-1.50	\\
\hline
R & 0.7064 &	0.8098 &	0.9412 &	0.9544\\
$\sigma$ & 0.44	& 0.37 &	0.18 &	0.16 \\
\hline
\end{tabular}
\caption[]{The four different types of post-DFT interaction energies calculated with the Hirshfeld-I method and the B3LYP functional. All values are in kcal/mol.}
\end{table}

\newpage
%Table 3 PBE+D
\begin{table}[ht!]
	\centering
		\begin{tabular}{lcccc}
		\hline
		Complex & C$_{6}^{iso}$&C$_{6}^{aniso}$&	full&	mixed \\
		\hline
He-He	&			-0.09	&	-0.09	&	-0.11	&	-0.11	\\
He-Ne	&			-0.13	&	-0.13	&	-0.17	&	-0.17	\\
He-Ar	&			-0.15	&	-0.15	&	-0.20	&	-0.20	\\
Ne-Ne	&			-0.19	&	-0.19	&	-0.25	&	-0.25	\\
Ne-Ar	&			-0.24	&	-0.24	&	-0.35	&	-0.35	\\
Ar-Ar	&			-0.38	&	-0.38	&	-0.64	&	-0.64	\\
L-He-N2	&			-0.14	&	-0.13	&	-0.18	&	-0.18	\\
T-He-N2	&			-0.17	&	-0.17	&	-0.24	&	-0.24	\\
He-FCl	&			-0.23	&	-0.24	&	-0.33	&	-0.33	\\
FCl-He	&			-0.31	&	-0.31	&	-0.37	&	-0.37	\\
Ne-CH$_{4}$	&			-0.35	&	-0.35	&	-0.49	&	-0.49	\\
CH$_{4}$-C$_{2}$H$_{4}$	&			-0.48	&	-0.54	&	-0.66	&	-0.71	\\
CH$_{4}$-CH$_{4}$	&			-0.59	&	-0.57	&	-0.85	&	-0.82	\\
SiH$_{4}$-CH$_{4}$		&		-0.92	&	-0.90	&	-1.35	&	-1.34	\\
C$_{2}$H$_{2}$-C$_{2}$H$_{2}$	&			-1.58	&	-1.57	&	-2.25	&	-2.24	\\
CO$_{2}$-CO$_{2}$	&			-1.15	&	-1.21	&	-1.96	&	-2.02	\\
OCS-OCS	&			-1.19	&	-1.33	&	-2.44	&	-2.59	\\
C$_{2}$H$_{4}$-C$_{2}$H$_{4}$	&			-1.54	&	-1.40	&	-2.50	&	-2.36	\\
\hline
R & 0.9595 &	0.9777 &	0.9902 &	0.9953 \\
$\sigma$ & 0.18	& 0.13 &	0.12 &	0.09 \\
\hline
\end{tabular}
\caption[]{The four different types of post-DFT interaction energies calculated with the Hirshfeld-I method and the PBE functional. All values are in kcal/mol.}
\end{table}

\newpage
%Table 4 TPSS+D
\begin{table}[ht!]
	\centering
		\begin{tabular}{lcccc}
		\hline
		Complex & C$_{6}^{iso}$&C$_{6}^{aniso}$&	full&	mixed \\
		\hline
He-He	&			-0.07	&	-0.07	&	-0.09	&	-0.09	\\
He-Ne	&			-0.11	&	-0.11	&	-0.14	&	-0.14	\\
He-Ar	&			-0.11	&	-0.11	&	-0.16	&	-0.16	\\
Ne-Ne	&			-0.14	&	-0.14	&	-0.20	&	-0.20	\\
Ne-Ar	&			-0.17	&	-0.17	&	-0.28	&	-0.28	\\
Ar-Ar	&			-0.24	&	-0.24	&	-0.49	&	-0.49	\\
L-He-N2	&			-0.10	&	-0.09	&	-0.14	&	-0.13	\\
T-He-N2	&			-0.12	&	-0.12	&	-0.18	&	-0.18	\\
He-FCl	&			-0.16	&	-0.16	&	-0.25	&	-0.25	\\
FCl-He	&			-0.16	&	-0.16	&	-0.22	&	-0.22	\\
Ne-CH$_{4}$	&			-0.20	&	-0.19	&	-0.33	&	-0.33	\\
CH$_{4}$-C$_{2}$H$_{4}$	&			-0.21	&	-0.27	&	-0.37	&	-0.43	\\
CH$_{4}$-CH$_{4}$	&			-0.27	&	-0.25	&	-0.51	&	-0.48	\\
SiH$_{4}$-CH$_{4}$	&			-0.43	&	-0.42	&	-0.83	&	-0.82	\\
C$_{2}$H$_{2}$-C$_{2}$H$_{2}$	&			-1.19	&	-1.18	&	-1.83	&	-1.82	\\
CO$_{2}$-CO$_{2}$	&			-0.61	&	-0.67	&	-1.38	&	-1.45	\\
OCS-OCS	&			-0.50	&	-0.65	&	-1.70	&	-1.85	\\
C$_{2}$H$_{4}$-C$_{2}$H$_{4}$	&			-0.79	&	-0.65	&	-1.72	&	-1.58	\\
\hline
R & 0.8572 &	0.8934 &	0.9791 &	0.9853 \\
$\sigma$ & 0.32	& 0.28 &	0.13 &	0.11 \\
\hline
\end{tabular}
\caption[]{The four different types of post-DFT interaction energies calculated with the Hirshfeld-I method and the TPSS functional. All values are in kcal/mol.}
\end{table}

\newpage
\section{Figures}
\begin{figure}[ht]
	\centering
\resizebox{1.0\textwidth}{!}{\includegraphics{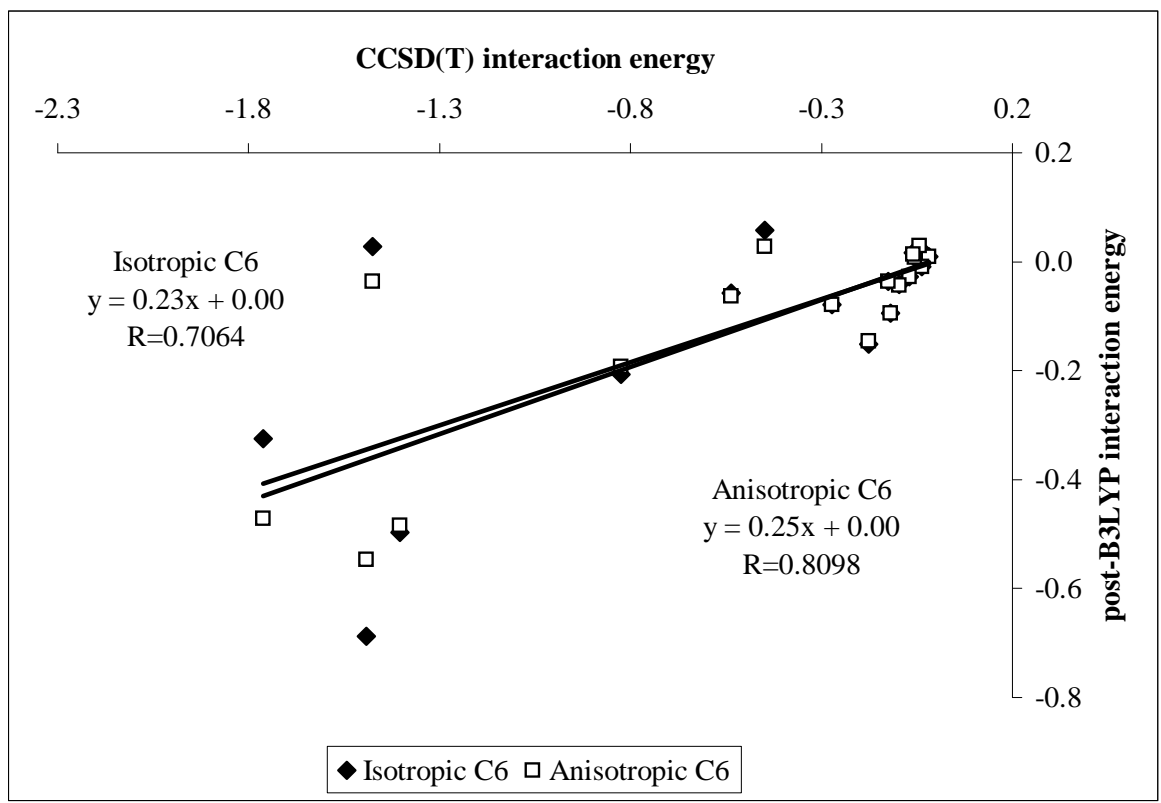}}
	\caption[]{Correlation between high level interaction energies and post-B3LYP interaction energies obtained with the isotropic and anisotropic C$_{6}$ models. All values are in kcal/mol.}
\end{figure}
\begin{figure}[ht]
	\centering
\resizebox{1.0\textwidth}{!}{\includegraphics{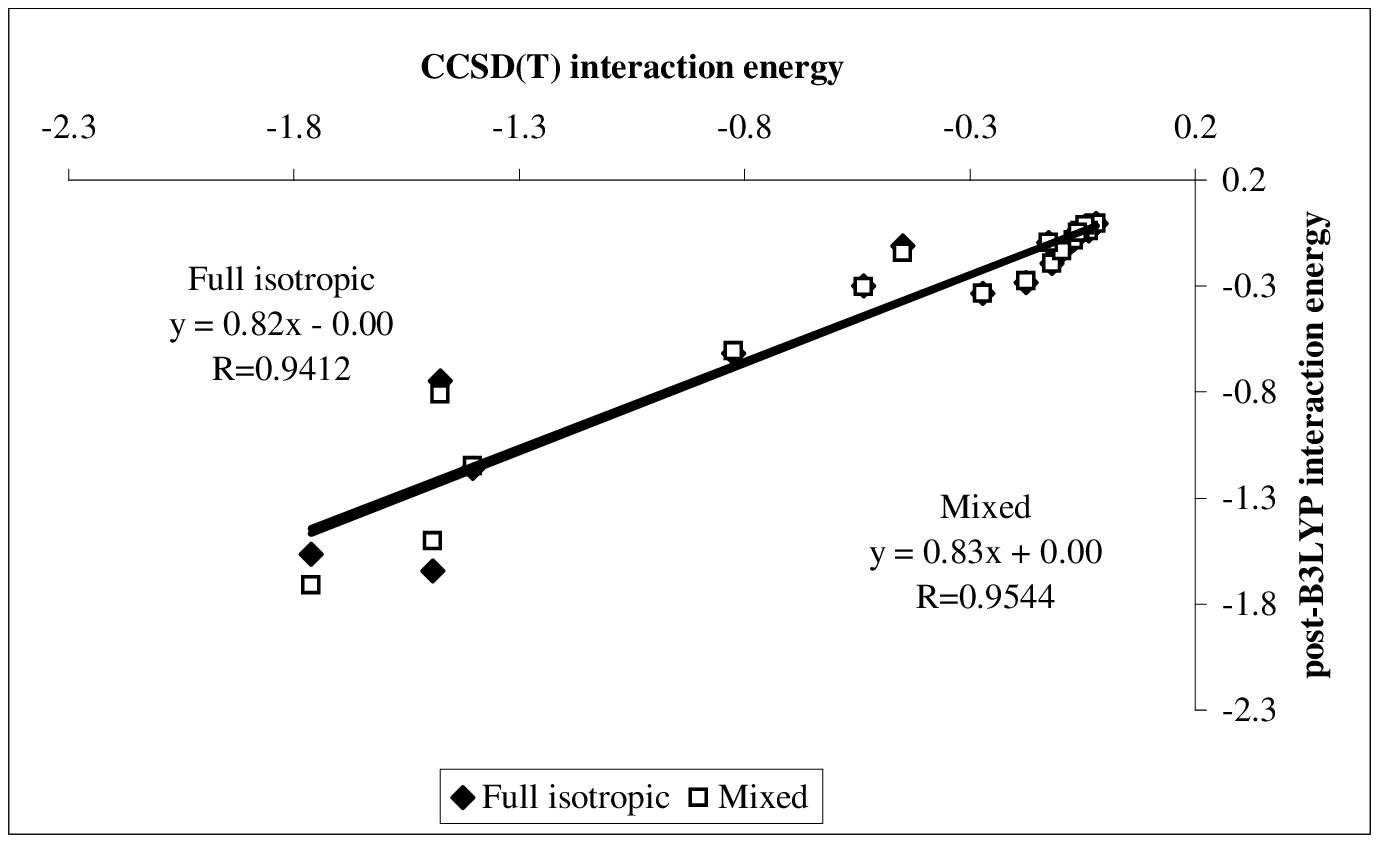}}
	\caption[]{Correlation between high level interaction energies and post-B3LYP interaction energies obtained with the full isotropic and mixed models. All values are in kcal/mol.}
\end{figure}
\begin{figure}[ht]
	\centering
\resizebox{1.0\textwidth}{!}{\includegraphics{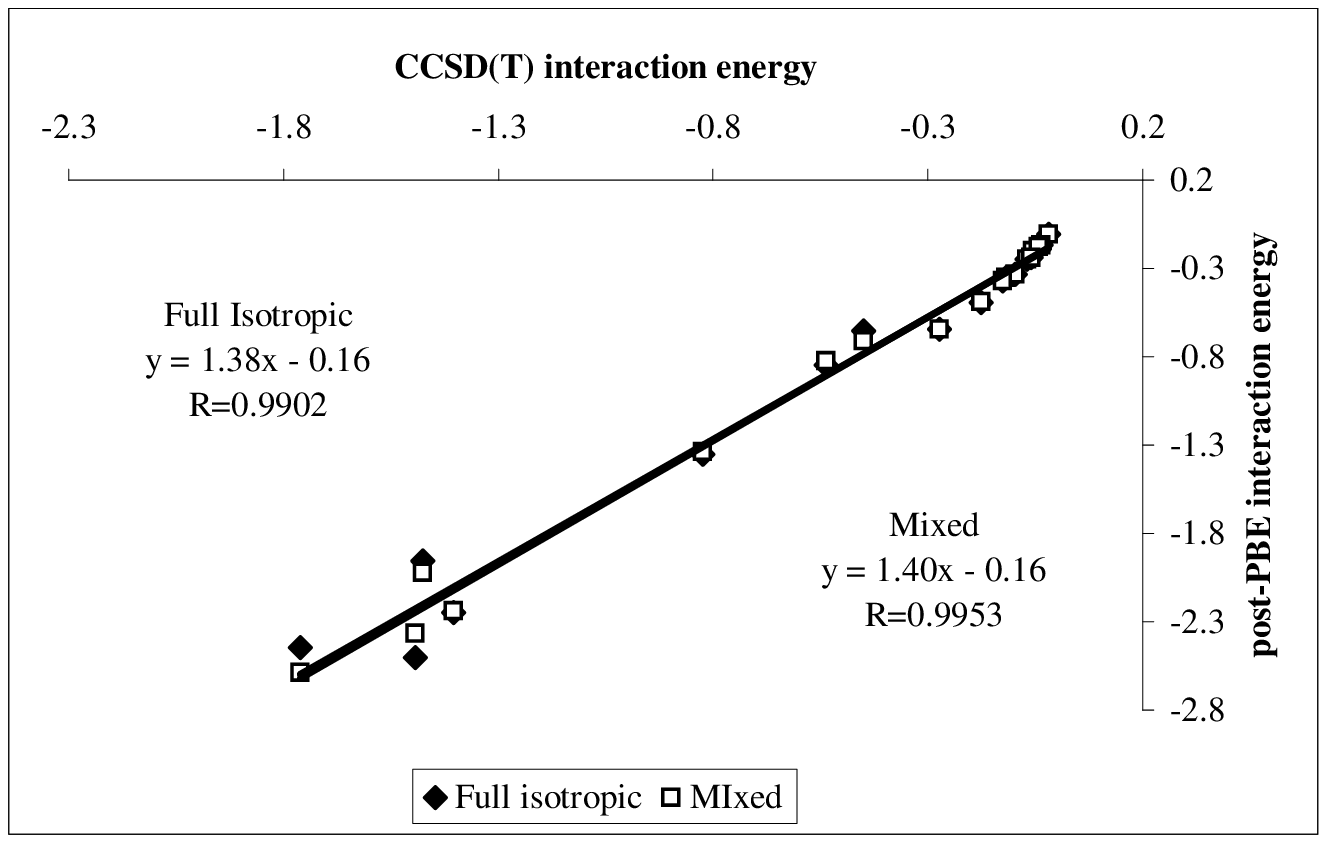}}
	\caption[]{Correlation between high level interaction energies and post-PBE interaction energies obtained with the full isotropic and mixed models. All values are in kcal/mol.}
\end{figure}
\begin{figure}[ht]
	\centering
\resizebox{1.0\textwidth}{!}{\includegraphics{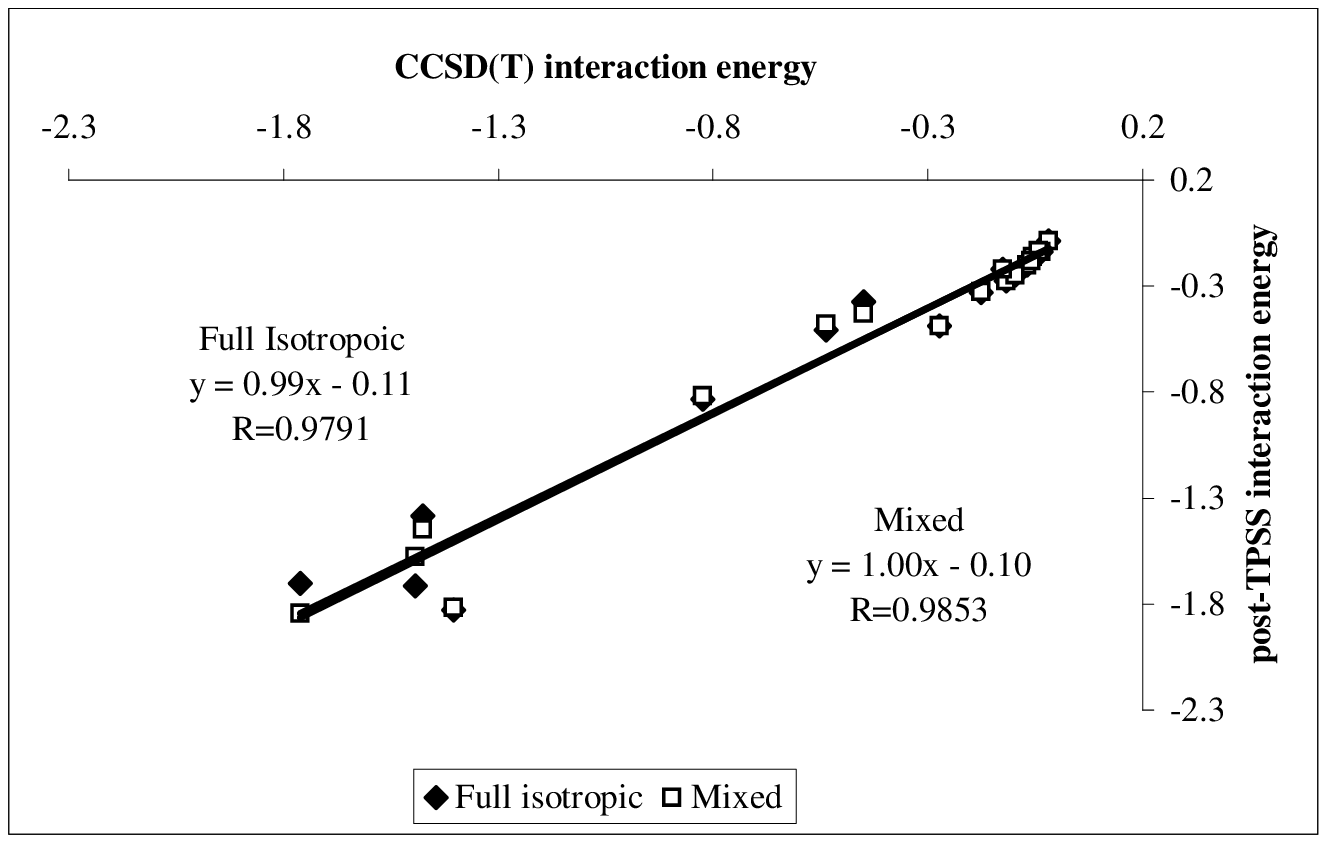}}
	\caption[]{Correlation between high level interaction energies and post-TPSS interaction energies obtained with the full isotropic and mixed models. All values are in kcal/mol.}
\end{figure}


\newpage
\begin{thebibliography}{99}
\bibitem{ref:Koch} W. Koch and M. C. Holthause, \textit{A Chemist's Guide to Density Functional Theory}, Willey-VCH, Weinheim, 2000.
%\bibitem{ref:Koch} Koch, W.; Holthausen, M. C.; \textit{A Chemist's Guide to Density Functional Theory}, Willey-VCH, Weinheim 2000.
\bibitem{ref:HobzaPCCP} P. Hobza, Phys. Chem. Chem. Phys. \textbf{10}, 2581 (2008).
%\bibitem{ref:HobzaPCCP} Hobza, P.; \textit{Phys. Chem. Chem. Phys.} \textbf{2008}, \textit{10}, 2581.
\bibitem{ref:Hobza} P. Hobza and J. \v{S}poner, Chem. Rev. (Washington, D. C.) \textbf{99}, 3247 (1999).
%\bibitem{ref:Hobza}Hobza, P.; \v{S}poner, J.; \textit{Chem. Rev. (Washington, D.C.)} \textbf{1999}, \textit{99}, 3247.
\bibitem{ref:Sinnokrot} M. O. Sinnokrot, E. F. Valeev and C. D. Sherrill, J. Am. Chem. Soc. \textbf{124}, 10887 (2002).
%\bibitem{ref:Sinnokrot} Sinnokrot, M.O.; Valeev, E.F.; Sherrill, C.D.; \textit{J. Am. Chem. Soc.} \textbf{2002}, \textit{124}, 10887.
\bibitem{ref:Versees} W. Vers\'{ee}s, S. Loverix, A. Vandemeulebroucke, P. Geerlings and J. Steyaert, J. Mol. Biol \textbf{338}, 1 (2004).
%\bibitem{ref:Versees} Vers\'ees, W.; Loverix, S.; Vandemeulebroucke, A.; Geerlings, P.; Steyaert, J.; \textit{J. Mol. Biol.} \textbf{2004}, \textit{338}, 1.
\bibitem{ref:Mignon1} P. Mignon, S. Loverix, F. De Proft and P. Geerlings, J. Phys. Chem. A \textbf{108}, 6038 (2004).
%\bibitem{ref:Mignon1} Mignon, P.; Loverix, S.; De Proft, F.; Geerlings, P.; \textit{J. Phys. Chem. A} \textbf{2004}, \textit{108}, 6038.
\bibitem{ref:Mignon2} P. Mignon, S. Loverix and P. Geerlings, Chem. Phys. Lett. \textbf{401}, 40 (2005).
%\bibitem{ref:Mignon2} Mignon, P.; Loverix, S.; Geerlings, P.; \textit{Chem. Phys. Lett.} \textbf{2005}, \textit{401}, 40.
\bibitem{ref:Mignon3} P. Mignon, S. Loverix, J. Steyaert and P. Geerlings, Nucleic. Acids Res. \textbf{33}, 1779 (2005).
%\bibitem{ref:Mignon3} Mignon, P.; Loverix, S.; Steyaert, J.; Geerlings, P.; \textit{Nucleic Acids Res..} \textbf{2005}, \textit{33}, 1779.
\bibitem{ref:Xu} X. Xu and W. A. Goddard III, Proc. Natl. Acad. Sci. U.S.A. \textbf{101}, 267 (2004).
%\bibitem{ref:Xu} Xu, X.; Goddard III, W. A.; \textit{Proc. Natl. Acad. Sci. U.S.A.} \textbf{2004}, \textit{101}, 267.
\bibitem{ref:Zhao1} Y. Zhao and D. G. Truhlar, J. Chem. Theory Comput. \textbf{1}, 415 (2005).
%\bibitem{ref:Zhao1} Zhao, Y.; Truhlar, D. G.; \textit{J. Chem. Thoer. Comput.} \textbf{2005}, \textit{1}, 415.
\bibitem{ref:Zhao2} Y. Zhao and D. G. Truhlar, J. Chem. Theory Comput. \textbf{3}, 289 (2007).
%\bibitem{ref:Zhao2} Zhao, Y.; Truhlar, D. G.; \textit{J. Chem. Thoer. Comput.} \textbf{2007}, \textit{3}, 289.
\bibitem{ref:Zhao3} Y. Zhao and D. G. Truhlar, Accounts of Chemical Research \textbf{41}, 157 (2008).
%\bibitem{ref:Zhao3} Zhao, Y.; Truhlar, D. G.; \textit{Accounts of Chemical Research} \textbf{2008}, \textit{41}, 157.
\bibitem{ref:Lilienfeld1} O. A. von Lillenfeld, I. Tavernelli, U. Rothlisberger and D. Sebastiani, Phys. Rev. Lett. \textbf{93}, 153004 (2004).
\bibitem{ref:Lilienfeld2} O. A. von Lillenfeld, I. Tavernelli, U. Rothlisberger and D. Sebastiani, Phys. Rev. B \textbf{71}, 195119 (2005).
\bibitem{ref:Lin1} I. C. Lin and U. Rothlisberger, CHIMIA \textbf{62}, 231 (2008).
%\bibitem{ref:Lin1} Lin, I. C.; Rothlisberger, U.; \textit{CHIMIA} \textbf{2008}, \textit{62}, 231.
\bibitem{ref:Lin2} I. C. Lin and U. Rothlisberger, Phys. Chem. Chem. Phys. \textbf{10}, 2730 (2008).
%\bibitem{ref:Lin2} Lin, I. C.; Rothlisberger, U.; \textit{Phys. Chem. Chem. Phys} \textbf{2008}, \textit{10}, 2730.
\bibitem{ref:Dion} M. Dion, H. Rydberg E. Schr\"{o}der, D. C. Langreth and B. I. Lundqvist, Phys. Rev. Lett \textbf{92}, 246401, (2004).
%\bibitem{ref:Dion} Dion, M.; Rydberg, H.; Schr\:{o}der, E.; Langreth, D. C.; Lundqvist, B. I.; \textit{Phys. Rev. Lett.} \textbf{2004}, \textit{92}, 246401.
\bibitem{ref:Ikura} H. Ikura, T. Tsuneda, T. Yanai and K. Hirao, J. Chem. Phys. \textbf{115}, 3540 (2001).
%\bibitem{ref:Ikura} Ikura, H.; Tsuneda, T.; Yanai, T.; Hirao, K.; \textit{J. Chem. Phys.} \textbf{2001}, \textit{115}, 3540.
\bibitem{ref:Sato} T. Sato, T. Tsuneda and K. Hirao, J. Chem. Phys. \textbf{126}, 234114 (2007).
%\bibitem{ref:Sato} Sato, T.; Tsuneda, T.; Hirao, K.; \textit{J. Chem. Phys.} \textbf{2007}, \textit{126}, 234114.
\bibitem{ref:Kamiya} M. Kamiya, T. Tsuneda and K. Hirao, J. Chem. Phys. \textbf{117}, 6010 (2002).
%\bibitem{ref:Kamiya} Kamiya, M.; Tsuneda, T.; Hirao, K.; \textit{J. Chem. Phys.} \textbf{2002}, \textit{117}, 6010.
\bibitem{ref:Grimme1} S. Grimme, J. Comput. Chem. \textbf{25}, 1463 (2004).
%\bibitem{ref:Grimme1} Grimme, S.; \textit{J. Comput. Chem.} \textbf{2004}, \textit{25}, 1463.
\bibitem{ref:Grimme2} S. Grimme, J. Comput. Chem. \textbf{27}, 1787 (2006).
%\bibitem{ref:Grimme2} Grimme, S.; \textit{J. Comput. Chem.} \textbf{2006}, \textit{27}, 1787.
\bibitem{ref:Grimme3} J. Anthony and S. Grimme, Phys. Chem. Chem. Phys. \textbf{8}, 5287 (2006).
%\bibitem{ref:Grimme3} Anthony, J.; Grimme, S.; \textit{Phys. Chem. Chem. Phys} \textbf{2006}, \textit{8}, 5287.
\bibitem{ref:Grimme4} S. Grimme, J. Anthony, T. Schwabe and C. M\"{u}ck-Lichtenfeld, Org. Biomol. Chem. \textbf{5}, 741 (2007).
%\bibitem{ref:Grimme4} Grimme, S.; Anthony, J.; Schwabe, T.; M\:{u}ck-Lichtenfeld, C.; \textit{Org. Biomol. Chem.} \textbf{2007}, \textit{5}, 741.
\bibitem{ref:Grimme5} S. Grimme, Angew. Chem. Int. Ed. \textbf{47}, 3430 (2008).
%\bibitem{ref:Grimme5} Grimme, S.; \textit{Angew. Chem. Int. Ed.} \textbf{2008}, \textit{47}, 3430.
\bibitem{ref:Jurecka} P. Jur\v{c}ka, J. \v{C}ern\'y, P. Hobza and D. R. Salahub, J. Comput. Chem. \textbf{28}, 555 (2007).
%\bibitem{ref:Jurecka} Jure\v{c}ka, P.; \v{C}ern\'y, J.; Hobza, P.; Salahub, D. R.; \textit{J. Comput. Chem.} \textbf{2007}, \textit{28}, 555.
\bibitem{ref:Cerny} J. \v{C}ern\'y and P. Hobza, Phys. Chem. Chem. Phys. \textbf{9}, 5291 (2007).
%\bibitem{ref:Cerny} \v{C}ern\'y, J.; Hobza, P.; \textit{Phys. Chem. Chem. Phys.} \textbf{2007}, \textit{9}, 5291.
\bibitem{ref:Chai} J. D. Chai and M. Head-Gorden, Phys. Chem. Chem. Phys. \textbf{10}, 6615 (2008).
%\bibitem{ref:Chai} Chai, J. D.; Head-Gorden, M.; \textit{Phys. Chem. Chem. Phys.} \textbf{2008}, \textit{10}, 6615.
\bibitem{ref:VanGisbergen1} S. J. A. Van Gisbergen, J. G. Snijders and E. J. Baerends, J. Chem. Phys. \textbf{103}, 9347 (1995).
%\bibitem{ref:VanGisbergen1} Van Gisbergen, S. J. A.; Snijders, J. G.; Baerends, E. J.; \textit{J. Chem. Phys.} \textbf{1995}, \textit{103}, 9347.
\bibitem{ref:VanGisbergen2} V. P. Osinga, S. J. A. Van Gisbergen, J. G. Snijders and E. J. Baerends, J. Chem. Phys \textbf{106}, 5091 (1997).
%\bibitem{ref:VanGisbergen2} Osinga, V. P.; Van Gisbergen, S. J. A.; Snijders, J. G.; Baerends, E. J.; \textit{J. Chem. Phys.} \textbf{1997}, \textit{106}, 5091. 
\bibitem{ref:BJ1} A. D. Becke and E. R. Johnson, J. Chem. Phys. \textbf{122}, 154104 (2005).
%\bibitem{ref:BJ1} Becke, A.D.; Johnson, E.R.; \textit{J. Chem. Phys.} \textbf{2005}, \textit{122}, 154104.  
\bibitem{ref:BJ2} E. R. Johnson and A. D. Becke, J. Chem. Phys. \textbf{123}, 024101 (2005).
%\bibitem{ref:BJ2} Johnson, E.R.; Becke, A.D.; \textit{J. Chem. Phys.} \textbf{2005}, \textit{123}, 024101.
\bibitem{ref:BJ3} A. D. Becke and E. R. Johnson, J. Chem. Phys. \textbf{123}, 154101, (2005).
%\bibitem{ref:BJ3} Becke, A.D.; Johnson, E.R.; \textit{J. Chem. Phys.} \textbf{2005}, \textit{123}, 154101. 
\bibitem{ref:BJ4} E. R. Johnson and A. D. Becke, J. Chem. Phys. \textbf{124}, 174104, (2006).
%\bibitem{ref:BJ4} Johnson, E.R.; Becke, A.D.; \textit{J. Chem. Phys.} \textbf{2006}, \textit{124}, 174104.
\bibitem{ref:BJ5} A. D. Becke and E. R. Johnson, J. Chem. Phys. \textbf{124}, 014104, (2006).
%\bibitem{ref:BJ5} Becke, A.D.; Johnson, E.R.; \textit{J. Chem. Phys.} \textbf{2006}, \textit{124}, 014104. 
\bibitem{ref:BJ6} A. D. Becke and E. R. Johnson, J. Chem. Phys. \textbf{127}, 154108 (2007).
%\bibitem{ref:BJ6} Becke, A.D.; Johnson, E.R.; \textit{J. Chem. Phys.} \textbf{2007}, \textit{127}, 154108. 
\bibitem{ref:disp1} A. Olasz, K. Vanommeslaeghe, A. Krishtal, T. Veszpr\'{e}mi, C. Van Alsenoy and P. Geerlings, J. Chem. Phys. \textbf{127}, 224105 (2007).
%\bibitem{ref:disp1} Olasz, A.; Vanommeslaeghe, K.; Krishtal, A.; Veszpr\'emi, T.; Van Alsenoy, C.; Geerlings, P.; \textit{J. Chem. Phys.} \textbf{2007}, \textit{127}, 224105.
\bibitem{ref:Hirshfeld} F. L. Hirshfeld, Theoret. Chem. Acta (Berl.) \textbf{44}, 129 (1977).
%\bibitem{ref:Hirshfeld}Hirshfeld, F. L. \textit{Theoret. Chim. Acta (Berl.)} \textbf{1977}, \textit{44}, 129.
\bibitem{ref:Rousseau} B. Rousseau, A. Peeters and C. Van Alsenoy, Chem. Phys. Lett. \textbf{324}, 189 (2002).
%\bibitem{ref:Rousseau}Rousseau, B.; Peeters, A.; Van Alsenoy, C.; \textit{Chem. Phys. Lett.}, \textbf{2000}, \textit{324}, 189.
\bibitem{ref:DeProft} F. De Proft, C. Van Alsenoy, A. Peeters, W. Langenaeker and P. Geerlings, J. Comp. Chem. \textbf{23}, 1198 (2002).
%\bibitem{ref:DeProft} De Proft, F.; Van Alsenoy, C.; Peeters, A.; Langenaeker, W.; Geerlings, P.; \textit{J. Comp. Chem.} \textbf{2002}, \textit{23}, 1198.
\bibitem{ref:polar} A. Krishtal, P. Senet, Y. Mingli and C. Van Alsenoy, J. Chem. Phys. \textbf{125}, 034312 (2006).
\bibitem{ref:B3LYP}  A. D. Becke, J. Chem. Phys. \textbf{98}, 5648 (1993). 
\bibitem{ref:PBE1} J. P. Perdew, K. Burke and M. Ernzerhof, Phys. Rev. Lett. \textbf{77}, 3865 (1996).
\bibitem{ref:PBE2} J. P. Perdew, K. Burke and M. Ernzerhof, Phys. Rev. Lett. \textbf{78}, 1396 (1997). 
\bibitem{ref:TPSS} J. M. Tao, J. P. Perdew, V. N. Staroverov and G. E. Scuseria, Phys. Rev. Lett. \textbf{91}, 146401 (2003).
%\bibitem{ref:polar} Krishtal, A.; Senet, P.; Mingli, Y.; Van Alsenoy, C. \textit{J. Chem. Phys.} \textbf{2006}, \textit{125}, 034312.
\bibitem{ref:H-I} P. Bultinck, C. Van Alsenoy, P. W. Ayers and R. Carbo-Dorca, J. Chem. Phys. \textbf{126}, 144111 (2007).
%\bibitem{ref:H-I} Bultinck, P.; Van Alsenoy, C.; Ayers, P. W.; Carbo-Dorca, R. \textit{J. Chem. Phys.} \textbf{2007}, \textit{126}, 144111. 
\bibitem{ref:SulfonicAcids} A. Krishtal, P. Senet and C. Van Alsenoy, J. Chem. Theory Comput. \textbf{4}, 2122 (2008). 
%\bibitem{ref:SulfonicAcids} Krishtal, A.; Senet, P.; Van Alsenoy, C.; \textit{J. Chem. Theory Comput.} \textbf{2008}, \textit{xxx}, xxx. ASAP, DOI10.1021/ct800295h
\bibitem{ref:Buckingham} A. D. Buckingham, Adv. Chem. Phys. \textbf{12}, 107 (1967).
%\bibitem{ref:Buckingham} Buckingham, A. D. \textit{Adv. Chem. Phys.} \textbf{1967}, \textit{12}, 107.
\bibitem{ref:Buckingham2} A. D. Buckingham, Intermolecular Interactions: from Diatomics to Biopolymers, Willey, New York, p1, 1978.
%\bibitem{ref:Buckingham2} Buckingham, A. D., \textit{Intermolecular Interactions: from Diatomics to Biopolymers}, \textbf{1978}, Pullman, B. Editor, Willey, New York, p1.
\bibitem{ref:London} F. London, Trans. Farad. Soc. \textbf{33}, 8 (1937).
%\bibitem{ref:London} London, F. \textit{Trans. Farad. Soc.} \textbf{1937}, \textit{33}, 8.
\bibitem{ref:Unsold} A. Uns\"{o}ld, Z. Physik \textbf{43}, 563 (1927).
%\bibitem{ref:Unsold} Uns\"{o}ld, A. \textit{Z. Physik} \textbf{1927}, \textit{43}, 563.
\bibitem{ref:Pullman} B. Pullman and P. Claverie, J. Caillet, Proc. Natl. Acad. Sci. U.S.A. \textbf{55}, 904 (1966).
%\bibitem{ref:Pullman} Pullman, B.; Claverie, P.; Caillet, J.; \textit{Proc. Natl. Acad. Sci. (USA)} \textbf{1966}, \textit{55}, 904.
\bibitem{ref:Rijks} W. Rijks and P. E. S. Wormer, J. Chem. Phys. \textbf{90}, 6507 (1989).
%\bibitem{ref:Rijks} Rijks, W.; Wormer, P. E. S.; \textit{J. Chem. Phys.} \textbf{1989}, \textit{90}, 6507.
\bibitem{ref:bsse-Boys} S. F. Boys and F. Bernardi, Mol. Phys. \textbf{10}, 553 (1970).
%\bibitem{ref:bsse-Boys} Boys, S.F.; Bernardi, F. \textit{Mol. Phys.} \textbf{1970}, \textit{10}, 553.
\bibitem{ref:bsse-Simon} S. Simon, M. Duran and J. J. Dannenberg, J. Chem. Phys. \textbf{105}, 11024 (1996).
%\bibitem{ref:bsse-Simon} Simon, S.; Duran, M.; Dannenberg. J.J. \textit{J. Chem. Phys.} \textbf{1996}, \textit{105}, 11024.
\bibitem{ref:geom-atom} T. J. Geise and D. M. York, Int. J. Quantum Chem. \textbf{98}, 388 (2004).
%\bibitem{ref:geom-atom} Giese, T.J.; York, D.M.; \textit{Int. J. Quantum Chem.} \textbf{2004}, \textit{98}, 388. 
\bibitem{ref:geom-hen2} K. Patel, P. R. Butler, A. M. Ellis and M. D. Wheeler, J. Chem. Phys. \textbf{119}, 909 (2003).
%\bibitem{ref:geom-hen2} Patel, K.; Butler, P.R.; Ellis, A.M.; Wheeler, M.D.; \textit{J. Chem. Phys.} \textbf{2003}, \textit{119}, 909.  
\bibitem{ref:geom-hefcl} R. Prosmiti, C. Cunha, P. Villareal and G. Delgado-Barrio, J. Chem. Phys \textbf{119}, 4216 (2003).
%\bibitem{ref:geom-hefcl}Prosmiti, R.; Cunha, C.; Villarreal, P.; Delgado-Barrio, G.; \textit{J. Chem. Phys.} \textbf{2003}, \textit{119}, 4216.  
\bibitem{ref:geom-sih4ch4} E. R. Johnson and G. A. DiLabio, Chem. Phys. Lett. \textbf{397}, 314 (2004). 
%\bibitem{ref:geom-sih4ch4} Johnson, E.R.; DiLabio, G.A.; \textit{Chem. Phys. Lett.} \textbf{2004}, \textit{397}, 314.  
\bibitem{ref:geom-co2co2} S. Tsuzuki, T. Uchimaru, M. Mikami and K. Tanabe, J. Chem. Phys. \textbf{109}, 2169 (1998).
%\bibitem{ref:geom-co2co2} Tsuzuki, S.; Uchimaru, T.; Mikami, M.; Tanabe, K.; \textit{J. Chem. Phys.} \textbf{1998}, \textit{109}, 2169.  
\bibitem{ref:geom-ocsocs} R. G. A. Bone, Chem. Phys. Lett. \textbf{206}, 260 (1993).
%\bibitem{ref:geom-ocsocs} Bone, R.G.A.; \textit{Chem. Phys. Lett.} \textbf{1993}, \textit{206}, 260.
\bibitem{ref:Sinnokrot2} M. O. Sinnokrot and C. D. Sherrill, J. Phys. Chem. A \textbf{108}, 10200 (2004).
\bibitem{ref:Helgaker1}A. Halkier, T. Helgaker, P. J\o rgensen, W. Klopper, H. Koch, J. Olsen and A. K. Wilson, Chem. Phys. Lett. \textbf{286}, 243, (1998).
\bibitem{ref:Helgaker2} K. L. Bak, P. J\o rgensen, J. Olsen, T. Helgaker and W. Klopper, J. Chem. Phys. \textbf{112}, 9229, (2000).

\bibitem{Gaussian}
 M. J. Frisch, G. W. Trucks, H. B. Schlegel, G. E. Scuseria, M. A. Robb, J. R. Cheeseman, J. A. Montgomery, Jr., T. Vreven, K. N. Kudin, J. C. Burant, J. M. Millam, S. S. Iyengar, J. Tomasi, V. Barone, B. Mennucci, M. Cossi, G. Scalmani, N. Rega, G. A. Petersson, H. Nakatsuji, M. Hada, M. Ehara, K. Toyota, R. Fukuda, J. Hasegawa, M. Ishida, T. Nakajima, Y. Honda, O. Kitao, H. Nakai, M. Klene, X. Li, J. E. Knox, H. P. Hratchian, J. B. Cross, V. Bakken, C. Adamo, J. Jaramillo, R. Gomperts, R. E. Stratmann, O. Yazyev, A. J. Austin, R. Cammi, C. Pomelli, J. W. Ochterski, P. Y. Ayala, K. Morokuma, G. A. Voth, P. Salvador, J. J. Dannenberg, V. G. Zakrzewski, S. Dapprich, A. D. Daniels, M. C. Strain, O. Farkas, D. K. Malick, A. D. Rabuck, K. Raghavachari, J. B. Foresman, J. V. Ortiz, Q. Cui, A. G. Baboul, S. Clifford, J. Cioslowski, B. B. Stefanov, G. Liu, A. Liashenko, P. Piskorz, I. Komaromi, R. L. Martin, D. J. Fox, T. Keith, M. A. Al-Laham, C. Y. Peng, A. Nanayakkara, M. Challacombe, P. M. W. Gill, B. Johnson, W. Chen, M. W. Wong, C. Gonzalez, and J. A. Pople, \textit{Gaussian 03}, Revision C.02 (Gaussian, Inc., Wallingford CT, 2004)
%\bibitem{Gaussian} Frisch, M. J.; Trucks, G. W.; Schlegel, H. B.; Scuseria, G. E.; Robb, M. A.; Cheeseman, J. R.; Montgomery, J. A.; Vreven, T. Jr.; Kudin, K. N.; Burant, J. C.; Millam, J. M.; Iyengar, S. S.; Tomasi, J.; Barone, V.; Mennucci, B.; Cossi, M.; Scalmani, G.; Rega, N.; Petersson, G. A.; Nakatsuji, H.; Hada, M.; Ehara, M.; Toyota, K.; Fukuda, R.; Hasegawa, J.; Ishida, M.; Nakajima, T.; Honda, Y.; Kitao, O.; Nakai, H.; Klene, M.; Li, X.; Knox, J. E.; Hratchian, H. P.; Cross, J. B.; Adamo, C.; Jaramillo, J.; Gomperts, R.; Stratmann, R. E.; Yazyev, O.; Austin, A. J.; Cammi, R.; Pomelli, C.; Ochterski, J. W.; Ayala, P. Y.; Morokuma, K.; Voth, G. A.; Salvador, P.; Dannenberg, J. J.; Zakrzewski, V. G.; Dapprich, S.; Daniels, A. D.; Strain, M. C.; Farkas, O.; Malick, D. K.; Rabuck, A. D.; Raghavachari, K.; Foresman, J. B.; Ortiz, J. V.; Cui, Q.; Baboul, A. G.; Clifford, S.; Cioslowski, J.; Stefanov, B. B.; Liu, G.; Liashenko, A.; Piskorz, P.; Komaromi, I.; Martin, R. L.; Fox, D. J.; Keith, T.; Al-Laham, M. A.; Peng, C. Y.; Nanayakkara, A.; Challacombe, M.; Gill, P. M. W.; Johnson, B.; Chen, W.; Wong, M. W.; Gonzalez C.; Pople, J. A. \textit{Gaussian 03}, Revision B.05 (Gaussian, Inc., Pittsburgh PA, 2003)
\end{thebibliography}
\end{document}